\documentstyle[12pt,twoside,fleqn,espcrc1,epsf]{article}
\newcommand{\la}{\langle}
\newcommand{\ra}{\rangle}

\newcommand{\beq}{\begin{eqnarray}}
\newcommand{\eeq}{\end{eqnarray}}

\newcommand{\btem}{\bibitem}

\newcommand{\AmS}{{\protect\the\textfont2
  A\kern-.1667em\lower.5ex\hbox{M}\kern-.125emS}}
\title{Pre-critical Chiral Fluctuations in Nuclear Medium --- precursors
 of chiral restoration at $\rho_{B}\not=0$ ---
\thanks{An expanded version of this work is see in; T. Hatsuda, T. Kunihiro
 and H. Shimizu, {\tt nuc-th/9810022}.}}

\author{T. Hatsuda\address{Physics Department,Kyoto University, Kyoto 606, 
Japan}
    and 
   T. Kunihiro\address{Faculty of Science and Technology,Ryukoku University, 
   Seta, Ohtsu 520-2194, Japan}}

\begin{document}
\maketitle

\begin{abstract}
It is shown that an enhancement  in the spectral function near 2$m_{\pi}$ 
threshold  in the $I$=$J$=0 channel  is a good  signal
 of the partial restoration of chiral symmetry in nuclear medium.
Several experiments are discussed to detect the enhancement,
which uses hadron-nucleus and photo-nucleus reactions.
\end{abstract}

\section{Introduction}
When  chiral symmetry is partially  restored in nuclear medium
\cite{HK94,BR}
one can expect a large fluctuation of the quark condensate
 $\la \bar {q}q\ra$, i.e., the order parameter of the chiral transition.
This implies \cite{HK94} that there arises a softening of a collective 
excitation 
 in  the scalar-isoscalar channel, leading to  
 (1)  partial degeneracy of the scalar-isoscalar particle
 (traditionally called the $\sigma$-meson) with the pion, and
 (2) decrease of the decay width of $\sigma$
  due to  the phase space suppression caused by (2) in the reaction
 $\sigma \rightarrow 2 \pi$:The significance of the
 $\sigma$ meson is discussed in \cite{HK94};see also \cite{kuni95}.
 
Although it is not a simple task to identify the $\sigma$ meson in free
 space\cite{pipi}, there will be a chance to see the elusive particle more
 clearly in nuclei where chiral symmetry might be partially restored.
Some experiments to produce the sigma meson in a nucleus was proposed
 in \cite{kuni95}.

One should, however, notice that describing the system with 
 the meson in terms of the meson mass and its width may become 
inadequate because of the strong interaction with the environmental particles.
Hence the proper observable to describes the system becomes the
spectral function.
Recently, a calculation of the spectral function in the sigma channel
has been performed with the $\sigma$-2$\pi$ coupling incorporated 
in the linear $\sigma$ model at finite $T$; it was shown that 
the enhancement of the spectral function in the $\sigma$-channel
just above the
two-pion threshold is the most distinct signal of the softening \cite{CH98}.

In this report, we shall show  that the spectral enhancement
associated with the partial chiral restoration  
takes place also at finite baryon density close to 
$\rho_0 = 0.17 {\rm fm}^{-3}$.

\section{Model calculation}
Before entering into the  explicit model-calculation, let us
 describe the general
 features of the spectral enhancement near the two-pion threshold.
 Consider the propagator 
 of the $\sigma$-meson at rest in the medium :
$D^{-1}_{\sigma} (\omega)= \omega^2 - m_{\sigma}^2 - $
$\Sigma_{\sigma}(\omega;\rho)$,
where $m_{\sigma}$ is the mass of $\sigma$ in the tree-level, and
$\Sigma_{\sigma}(\omega;\rho)$ is 
the loop corrections
 in the vacuum as well as in the medium.
 The corresponding spectral function is given by 
\beq
\rho_{\sigma}(\omega) = - \pi^{-1} {\rm Im} D_{\sigma}(\omega).
\eeq
Near the two-pion threshold, 
${\rm Im} \Sigma_{\sigma} \propto \theta(\omega - 2 m_{\pi}) \ 
	 \sqrt{1 - {4m_{\pi}^2 \over \omega^2}} $ in 
the one-loop order.
 When chiral symmetry is being restored,
 $m_{\sigma}^*$ (``effective mass'' of $\sigma$ 
 defined as a zero of the real part of the propagator
 ${\rm Re}D_{\sigma}^{-1}(\omega = m_{\sigma}^*)=0$)
  approaches to $ m_{\pi}$.  Therefore,
 there exists a density $\rho_c$ at which 
 ${\rm Re} D_{\sigma}^{-1}(\omega = 2m_{\pi})$
 vanishes even before the complete $\sigma$-$\pi$
 degeneracy takes place; namely
 ${\rm Re} D_{\sigma}^{-1} (\omega = 2 m_{\pi}) =
 [\omega^2 - m_{ \sigma}^2 -
 {\rm Re} \Sigma_{\sigma} ]_{\omega = 2 m_{\pi}} = 0$.
At this point, the spectral function can be  solely represented by the
 imaginary part of the self-energy;
\beq
\rho_{\sigma} (\omega \simeq  2 m_{\pi}) 
 =  - {1 \over \pi \ {\rm Im}\Sigma_{\sigma} }
 \propto {\theta(\omega - 2 m_{\pi}) 
 \over \sqrt{1-{4m_{\pi}^2 \over \omega^2}}},
\eeq
which shows an enhancement of the spectral function at the $2m_{\pi}$
threshold. We remark that this enhancement is generically 
 correlated with the  partial restoration of chiral symmetry.

To make the argument more quantitative,
let us evaluate $\rho_{\sigma}(\omega)$ in a {\em toy} model, namely
 the SU(2) linear $\sigma$-model:
\beq
\label{model-l}
{\cal L}  =  {1 \over 4} {\rm tr} [\partial M \partial M^{\dagger}
 - \mu^2 M M^{\dagger} 
  - {2 \lambda \over 4! } (M M^{\dagger})^2   - h (M+M^{\dagger}) ],
\eeq
where tr is for the flavor index and  
 $M = \sigma + i \vec{\tau}\cdot \vec{\pi}$.
 Although the model has only a limited number of parameters
 and is not a precise low energy representation of QCD, we 
emphasize that it does  describe the
 pion dynamics qualitatively well up to 1GeV as shown by
 Chan and Haymaker \cite{BW}, where the Pad$\acute{e}$ approximant is 
used for the scattering matrix.
 The coupling constants $\mu^2, \lambda$ and $h$ have
 been determined in the
 vacuum to reproduce $f_{\pi}=93$ MeV, $m_{\pi}=140$ MeV as well as
 the s-wave $\pi$-$\pi$ scattering phase shift in the one-loop order.

The nucleon sector and the interaction with the nucleon of the mesons 
are given by,
\beq
\label{int-nm}
{\cal L}_{I}(N, M) =
  - g \chi \bar{N} U_5 N  - m_0  \bar{N} U_5 N ,
\eeq
where we have used a polar representation 
 $ \sigma + i  \vec{\tau}\cdot \vec{\pi} \gamma_5 \equiv \chi U_5 $ 
 for convenience. 
 The first term in (\ref{int-nm}) with the coupling constant $g$
 is a standard chiral invariant coupling in the linear $\sigma$
 model. Although the second term with a new parameter $m_0$
 is   not usually taken into account in the literatures, 
 it is also chiral invariant and non-singular, so there is no
 compelling reason to dismiss it.

Under the dynamical breaking of chiral symmetry
 in the vacuum ($\langle \sigma \rangle_{\rm vac} \equiv \sigma_0 \neq 0$), 
 eq.(\ref{int-nm}) is expanded in
 terms of $\sigma/\sigma_0$ and $\vec{\pi}/\sigma_0$:
it is found that the standard constraint $g_s = g_p$ is relaxed
 without conflicting with chiral symmetry due to the term with $m_0$;
 the term proportional to $m_0 \pi^2$ appears to
 preserve chiral symmetry. 

 In the following, we treat the effect of the meson-loop as well as
 the baryon density  as a perturbation 
 to the vacuum quantities. Therefore, our loop-expansion  
  is valid only at relatively low
 densities.  The full self-consistent
 treatment of the problem requires  systematic resummation of loops
 similar to what was developed at finite $T$ \cite{CH98}. 

We parametrize the chiral condensate in nuclear matter
 $\langle \sigma \rangle$ as
$\langle \sigma \rangle \equiv  \sigma_0 \ \Phi(\rho)$.
In the linear density approximation,
 $\Phi(\rho) = 1 - C \rho / \rho_0 $ with
 $C = (g_{\rm s} /\sigma_0 m_{\sigma}^2) \rho_0$.
 Instead of using $g_{\rm s}$, we  use $\Phi$  as a basic parameter in the 
 following analysis.  The plausible value of $\Phi(\rho = \rho_0)$ is
 0.7 $\sim$ 0.9 \cite{HK94}.

\section{Results and discussions}

The spectral function together with ${\rm Re} D_{\sigma}^{-1}(\omega)$  
 calculated with a linear sigma model are shown 
  in Fig.1 and 2: The characteristic enhancements of the spectral
 function is seen just above the 2$m_{\pi}$.
It is also to be noted that even before
the $\sigma$-meson mass $m_{\sigma}^*$ and $m_{\pi}$ in the medium 
are degenerate,i.e., the chiral-restoring point, 
 a large enhancement
 of the spectral function near the $2m_{\pi}$ is seen.

\begin{figure}[htb]
\begin{minipage}[t]{75mm}
\epsfxsize=6.2cm
\centerline{\epsfbox{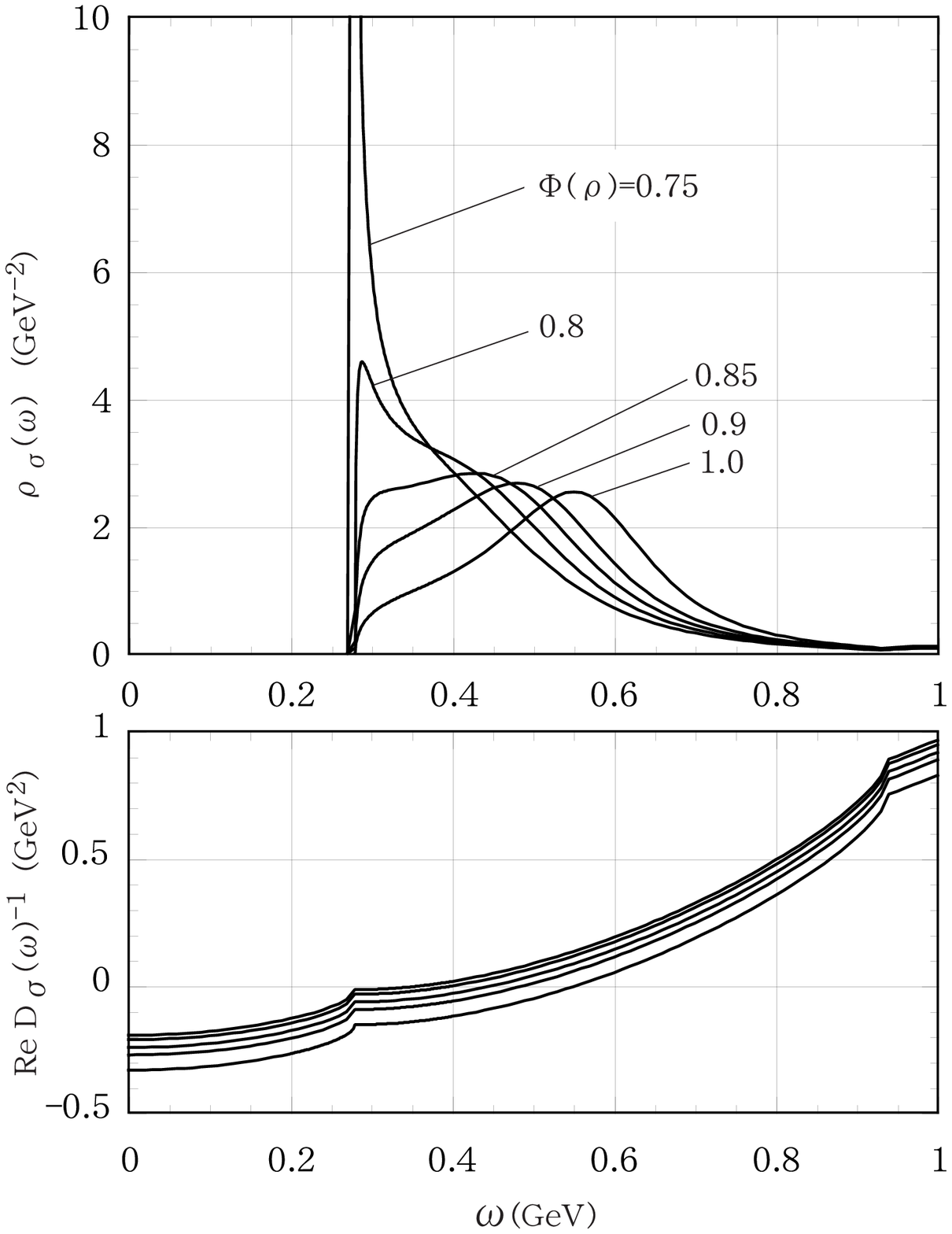}}
\caption{Spectral function for $\sigma$ and  the 
 real part of the inverse propagator for several values of
 $\Phi = \la \sigma \ra / \sigma_0$ with
 $m_{\sigma}^{peak} = 550$ MeV. In the lower panel,
 $\Phi$ decreases from bottom to top.}
\label{fig.1}
\end{minipage}
 %
\hspace{\fill}
\begin{minipage}[t]{76mm}
\epsfxsize=6.2cm
\centerline{\epsfbox{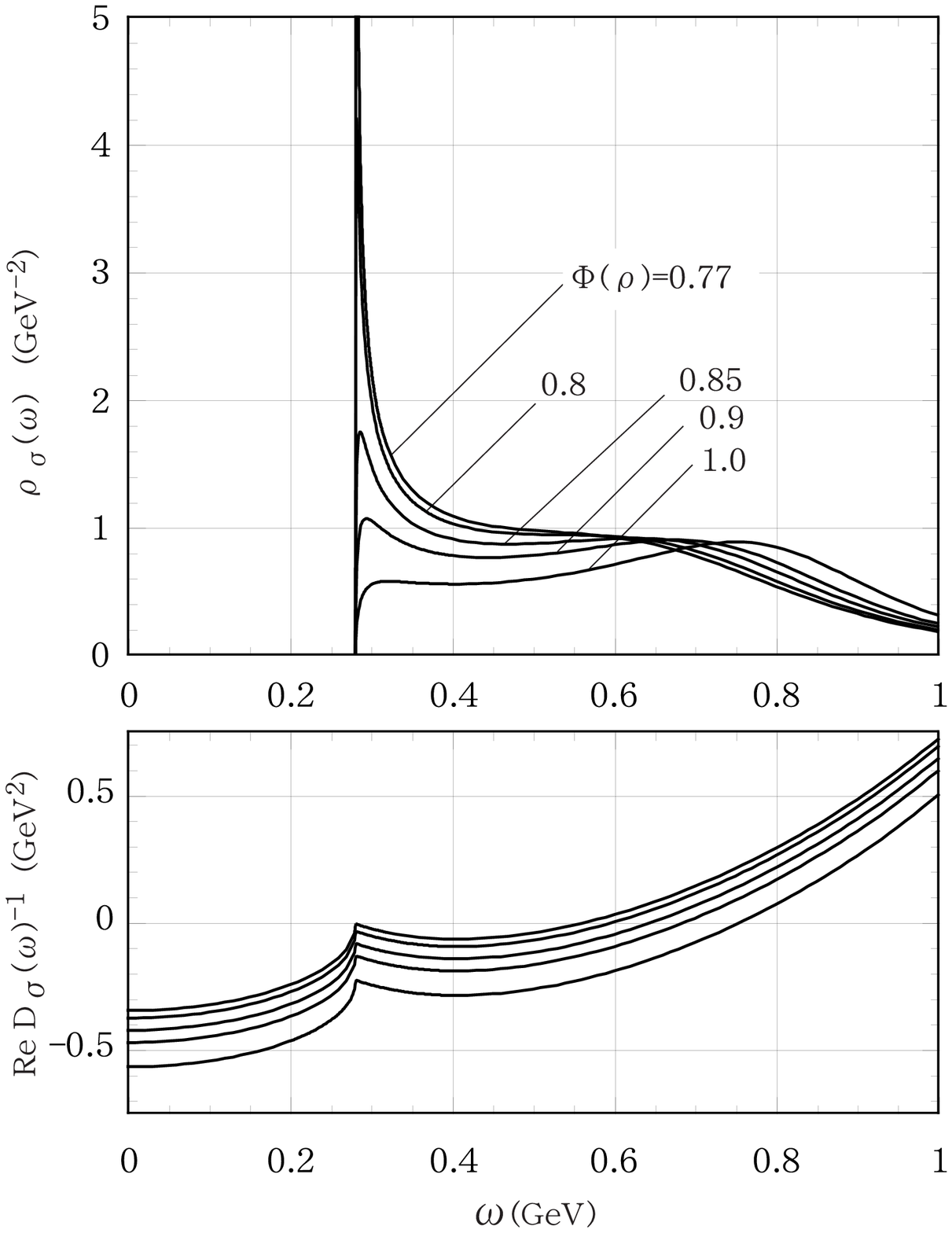}}
\caption{Same with Fig.1 for $m_{\sigma}^{peak}= 750$ MeV }
\label{fig.2}
\end{minipage}
\end{figure}

To confirm the threshold enhancement,
measuring 2$\pi^0$ and 
$2\gamma$ in experiments with hadron/photon beams off
 the  heavy nuclear targets are useful. 
 Measuring $\sigma \rightarrow 2 \pi^0 \rightarrow
  4\gamma$ is experimentally feasible 
 \cite{4gamma}, which is free from the $\rho$ meson background
  inherent in the $\pi^+\pi^-$ measurement.
 Measuring of 2 $\gamma$'s from the electromagnetic decay of the $\sigma$
 is interesting because of the small final state
 interactions, although the branching ratio is small.\footnote{
One needs also to fight with large 
 background of photons mainly coming from $\pi^0$s.}
 Nevertheless,  if the enhancement is prominent,
 there is a chance to find the signal.  
 When $\sigma$ has a finite three momentum,
one can detect dileptons  
 through the scalar-vector mixing in matter: $\sigma \to \gamma^* \to
 e^+ e^-$. 

Recently  CHAOS collaboration  \cite{ppp} measured the 
$\pi^{+}\pi^{\pm}$
invariant mass distribution $M^A_{\pi^{+}\pi^{\pm}}$ in the
 reaction $A(\pi^+, \pi^{+}\pi^{\pm})X$ with the 
 mass number $A$ ranging
 from 2 to 208: They observed that
the   yield for  $M^A_{\pi^{+}\pi^{-}}$ 
 near the 2$m_{\pi}$ threshold is close to zero 
for $A=2$, but increases dramatically with increasing $A$. They
identified that the $\pi^{+}\pi^{-}$ pairs in this range of
 $M^A_{\pi^{+}\pi^{-}}$ is in the $I=J=0$ state.
The $A$ dependence of the 
 the invariant mass distribution presented in \cite{ppp} 
 near 2$m_{\pi}$ threshold has a close
 resemblance to our model calculation in Fig.1 and 2, which suggests
 that this experiment may already provide
  a hint about how the partial restoration of chiral symmetry
 manifest itself at finite density.\footnote{See \cite{wambach} for
 other approaches to explain the CHAOS data.}

We remark that (d, $^3$He)  reactions is also useful to explore the
 spectral enhancement because of the
large incident flux.
 The incident kinetic energy $E$ of the  deuteron in the laboratory
 system is  estimated to be
  $1.1 {\rm GeV} < E < 10$ GeV, 
  to cover the spectral function 
 in the range  $2m_{\pi} < \omega < 750$ MeV.

To make the calculation more realistic, one needs
 to incorporate the two-loop  diagrams, which
 is expected, however, 
to hardly change the enhancement near the two-pi threshold 
discussed here.

In conclusion, we would like to express our sincere thanks to Prof. Yazaki
 for his interest in our work summarized in \cite{HK94} and 
the encouragement given to us for a long time.

\end{document}